\documentclass{article}
\usepackage{amssymb}

\input{tcilatex}
\begin{document}

\title{The Real Meaning of Complex Minkowski-Space World-Lines}
\date{11.17.2009}
\author{T.M. Adamo$^{1}$ \& E.T. Newman$^{2}$ \and $^{1}$University of
Oxford, Mathematical Institute, \and 24-29 St Giles, Oxford, OX1 3LB, UK
\and $^{2}$University of Pittsburgh, Department of Physics \& Astronomy,
\and Pittsburgh, PA 15213, USA}
\maketitle

\begin{abstract}
In connection with the study of shear-free null geodesics in Minkowski
space, we investigate the real geometric effects in real Minkowski space
that are induced by and associated with complex world-lines in complex
Minkowski space. It was already known, in a formal manner, that complex
analytic curves in complex Minkowski space induce shear-free null geodesic
congruences. \ Here we look at the direct geometric connections of the
complex line and the real structures. Among other items, we show, in
particular, how a complex world-line projects into the real Minkowski space
in the form of a real shear-free null geodesic congruence.
\end{abstract}

\section{Introduction}

It has been know for several years that \textit{regular} shear-free null
geodesic congruences (NGCs) in Minkowski space are generated, in a formal
manner, by four complex functions of a single complex variable, $\tau ,$
often designated by $\xi ^{a}(\tau )=(\xi ^{0}(\tau ),\xi ^{1}(\tau ),\xi
^{2}(\tau ),\xi ^{3}(\tau ))$\cite{PhysicalContent,Footprints}. It turns out
that these four functions can be naturally interpreted as determining a
complex world-line in complexified Minkowski space.\cite{Review} With
complex Minkowski coordinates, ($z^{a}$), the world-line is described by $%
z^{a}=\xi ^{a}(\tau ).$ It is the purpose of this note to try to flesh-out
this interpretation in detail and in particular to see what direct \textit{%
real} meaning can be assigned to these complex world-lines. (Though this
discussion could be extended to asymptotically flat space-times and
asymptotically shear-free NGC we will restrict the discussion to shear free
Null Geodesic Congruences in Minkowski space-time $\mathbb{M}$ and leave the
more general case for a later time.)

After the formal analytic discovery that regular shear-free NGCs could be
generated by the four analytic functions $\xi ^{a}(\tau ),$ the complex
Minkowski space interpretation arose from the following observation:
consider an arbitrary complex analytic world-line in complex Minkowski
space, $z^{a}=\xi ^{a}(\tau )$ and construct its future (surface forming) 
\textit{complex} light-cone. Many of the rays (the null geodesics) of this
complex cone `pierce' the real Minkowski space. \textit{\ }Of particular
importance is the intersection region of these complex light-cones with
complex null infinity, $\mathfrak{I}_{\mathbb{C}}^{+}.$ For each complex
cone this intersection (a complex `cut') is a complexified sphere.
Restricting these cuts to real null infinity, namely to $\mathfrak{I}^{+}$,
leads to real cuts, $S^{2}.$ (Note that when $\mathfrak{I}^{+}$ is
coordinatized by the standard Bondi coordinates ($u,\zeta ,\overline{\zeta }$%
), these real cuts can be given by $u=G_{R}(s,\zeta ,\overline{\zeta })$
with $s$ parametrizing the family of cuts.) The (\textit{twisting)
shear-free real} null geodesic congruences are completely determined by the
tangent directions, ($L,\overline{L}$), of these real cuts. The
determination of the real cuts forces a restriction on the range of the
complex parameter $\tau .$ The allowed values of $\tau $ are restricted to
the values $\tau =s+i\Lambda (s,\zeta ,\overline{\zeta })$, $s$ and $\Lambda
(s,\zeta ,\overline{\zeta })$ real, with $\Lambda (s,\zeta ,\overline{\zeta }%
)$ determined by the complex world-line.

\qquad In Sec. 2 we first outline our notation. This is then followed by a
method for the description in Minkowski space of arbitrary NGCs and the
specialization to shear-free NGCs. The relationships just mentioned, of the
complex cuts on $\mathfrak{I}_{\mathbb{C}}^{+}$ to the real shear-free NGCs,
is revisted in detail. In this context we review the procedure for the
restriction of these cuts to the real $\mathfrak{I}^{+}$ and the
determination of both the cut function $G_{R}(s,\zeta ,\overline{\zeta })$
and $\Lambda (s,\zeta ,\overline{\zeta })$.\ The section ends with a review
of the so-called "optical equations and parameters" \cite{Scholarpedia},
which play an important role in the later discussions.

In Section 3 we discuss some of the real geometric effects which emerge from
the study of the shear-free NGCs once one has taken into account the reality
constraint. \ We observe how the general complex world-line differs from a
pure real world-line once the reality constraints on the cut function $%
u=G(s,\zeta ,\overline{\zeta })$ have been used. \ In particular, we observe
that a complex-world line generates distorted 2-sphere cross-sections of $%
\mathfrak{I}^{+}.$ This is in contrast to the undistorted (spherical)
cross-sections generated by real world-lines. Further we note that the
imaginary part of a complex world-line is a measure of the twist of the
(real) NGC it describes. \ In addition, using both the tangent directions ($%
L,\overline{L}$) and an implicit relationship between the real parameter $s$
and the retarded time, $\sqrt{2}u=t-r$, we give an explicit Minkowski-space
coordinate description of the entire associated shear free NGC. \ The
section concludes with some remarks on how the reality constraints on the
complex cuts affect the CR structure of $\mathfrak{I}^{+}$ associated with
the shear-free NGCs.

The main results of this work are contained in Sec. 4 where we demonstrate
how one can map or project the complex worldline (with the reality condition
imposed on $\tau $) into the real Minkowski space. For each complex
world-line this mapping directly leads to a complete regular shear-free NGC.

In Section 5 the development of the caustics of shear-free NGCs (e.g., \cite%
{Mink}) is studied. This involves finding the space-time regions, in terms
of the Minkowski coordinates, where the optical parameter $\rho $, (i.e.,
the complex divergence of the shear-free NGC), "blows up." \ Generically the
caustic, at any fixed time, appears as a topologically $S^{1}$ "ring" moving
through the Minkowski space in time, i.e., \textit{a real closed `string'}.
\ The behavior of the caustic for certain non-generic cases is also
investigated. \ 

Section 6 contains a discussion of these results.

\section{Foundational Material}

\subsection{Shear-Free Null Geodesic Congruences in Minkowski space}

We begin with a description of the notation that is used. The coordinates of
real and complex Minkowski space are denoted by $x^{a}$ and $z^{a}$
respectively$.$ The Bondi coordinates on real real $\mathfrak{I}^{+}$ are $%
(u,\zeta ,\bar{\zeta}).$ $u$ labels slices or cuts while the complex
stereographic coordinates $(\zeta ,\bar{\zeta}),$ with $\zeta =e^{i\varphi
}\cot \frac{\theta }{2}$, label the individual null (generators) geodesics
of $\mathfrak{I}^{+}$. For the complexified $\mathfrak{I}^{+},$ i.e., $%
\mathfrak{I}_{\mathbb{C}}^{+}$ , $u$ takes on complex values and $\bar{\zeta}
$ $\rightarrow \widetilde{\zeta }$ is allowed complex values close to $\bar{%
\zeta}.$ \ At any space-time point one can introduce a null tetrad, $\{l,n,m,%
\bar{m}\}$ with standard inner product relations%
\begin{equation}
l^{a}n_{a}=-m^{a}\bar{m}_{a}=1,  \label{IP}
\end{equation}%
all other inner products vanishing. \ An arbitrary null geodesic starting at
any arbitrary point $x_{0}^{a}$ in the $l^{a}$ direction is given by%
\[
x^{a}=x_{0}^{a}+l^{a}r, 
\]%
with $r$ the affine parameter. We often make use of the explicit
representation for the tetrad $\{\hat{l},\hat{n},\hat{m},\overline{\hat{m}}%
\} $\ given by:%
\begin{eqnarray}
\hat{l}^{a} &=&\frac{\sqrt{2}}{2}\left( 1,\frac{\zeta +\bar{\zeta}}{1+\zeta 
\bar{\zeta}},-i\frac{\zeta -\bar{\zeta}}{1+\zeta \bar{\zeta}},\frac{-1+\zeta 
\bar{\zeta}}{1+\zeta \bar{\zeta}}\right) ,  \label{Stet} \\
\hat{n}^{a} &=&\frac{\sqrt{2}}{2}\left( 1,-\frac{\zeta +\bar{\zeta}}{1+\zeta 
\bar{\zeta}},i\frac{\zeta -\bar{\zeta}}{1+\zeta \bar{\zeta}},\frac{1-\zeta 
\bar{\zeta}}{1+\zeta \bar{\zeta}}\right) ,  \nonumber \\
\hat{m}^{a} &=&\frac{\sqrt{2}}{2}\left( 0,\frac{1-\bar{\zeta}^{2}}{1+\zeta 
\bar{\zeta}},-i\frac{1+\bar{\zeta}^{2}}{1+\zeta \bar{\zeta}},\frac{2\bar{%
\zeta}}{1+\zeta \bar{\zeta}}\right) ,  \nonumber
\end{eqnarray}%
that is allowed to `swing' around the entire sphere, $(\zeta ,\bar{\zeta})$ $%
\in S^{2}$, of null directions, at any space-time point.

Using these coordinates, tetrad and parameters, $(u,\zeta ,\bar{\zeta})\ $%
and appropriate choice of $L,$ any NGC, in Minkowski space, can be expressed
as \cite{Mink}:%
\begin{equation}
x^{a}=u\hat{t}^{a}-\bar{L}\hat{m}^{a}-L\overline{\hat{m}}^{a}+(r-r_{0})\hat{l%
}^{a},  \label{Trans}
\end{equation}%
where $L=L(u,\zeta ,\bar{\zeta})$ is an arbitrary holomorphic function, $%
r_{0}(u,\zeta ,\bar{\zeta})$ is the arbitrary origin of the affine parameter 
$r$, and 
\begin{equation}
\sqrt{2}\hat{t}^{a}=\hat{n}^{a}+\hat{l}^{a}=(\sqrt{2},0,0,0),  \label{t}
\end{equation}%
is a time-like vector. \ The parameters $(u,\zeta ,\bar{\zeta})$ which label
each null geodesic of the congruence are simply the Bondi coordinates of the
points of $\mathfrak{I}^{+}$intersected by the geodesic.

Equation (\ref{Trans}) can be interpreted in two ways: either as an
expression for the NGCs or as a coordinate transformation between the
standard Minkowski coordinates $x^{a}$ and the null geodesic coordinates $%
(u,r,\zeta ,\bar{\zeta})$. \ 

It is known \cite{Aronson} that the complex shear of such a congruence (\ref%
{Trans}) is given by:%
\begin{equation}
\sigma (u,\zeta ,\bar{\zeta})=\eth L+LL^{\cdot },  \label{Shear1}
\end{equation}%
where $\eth $ is the usual spin-weighted operator on the 2-sphere and $%
L^{\cdot }\equiv \partial _{u}L$. \ In the next subsection we return to $%
\sigma $ and its complimentary optical parameter, the complex divergence $%
\rho $.

As we are interested in \textit{shear-free} NGCs in $\mathbb{M}$, we look
for those NGCs given by (\ref{Trans}) where $\sigma (u,\zeta ,\bar{\zeta}%
)=0, $ i.e., where $L$ satisfies%
\begin{equation}
\eth L+LL^{\cdot }=0.  \label{Shear2}
\end{equation}

Solutions to the shear-free equation (\ref{Shear2}) are found by first
introducing a complex potential function $\tau =T(u,\zeta ,\bar{\zeta})$,
which satisfies the PDE:%
\begin{equation}
\eth T+LT^{\cdot }=0.  \label{CReq}
\end{equation}

Note that at this point we must consider the complexification of $\mathfrak{I%
}^{+}$ by allowing $u$ to take on complex values and free $\bar{\zeta}$ \
from being the complex conjugate of $\zeta .$ We also assume that $L$ is
complex analytic in the three independent arguments $(u,\zeta ,\bar{\zeta}).$

As explained later, Eq.(\ref{CReq}) is actually the CR equation that assigns
a CR structure to $\mathfrak{I}^{+}$ that is associated with any shear-free
NGC structure. For now it can just be viewed as a relation that the
potential function $\tau =T(u,\zeta ,\bar{\zeta})$ must satisfy if $%
L(u,\zeta ,\bar{\zeta})$ is known. \ Assuming that this potential function
can be inverted to give a complex "cut function" of $\mathfrak{I}_{\mathbb{C}%
}^{+}$%
\begin{equation}
u=G(\tau ,\zeta ,\bar{\zeta})\Leftrightarrow \tau =T(u,\zeta ,\bar{\zeta}),
\label{CutF}
\end{equation}%
replace $T(u,\zeta ,\bar{\zeta})$ by $G(\tau ,\zeta ,\bar{\zeta})$ as an
independent variable. \ Repeated implicit differentiations of $G$ transforms
the shear-free condition, Eqs.(\ref{Shear2}) and the expression for $L,$ (%
\ref{CReq}) to the relations \cite{Review,UCF}:%
\begin{eqnarray}
\eth _{(\tau )}^{2}G(\tau ,\zeta ,\bar{\zeta}) &=&0,  \label{GCE} \\
L(u,\zeta ,\bar{\zeta}) &=&\eth _{(\tau )}G.  \label{AngF}
\end{eqnarray}%
Here, $\eth _{(\tau )}$ indicates the application of the $\eth $-operator
while the variable $\tau $ is held constant. \ The $u$-dependence of $L$ is
recovered in equation (\ref{AngF}) by first applying $\eth _{(\tau )}$ to
the solution $u=G(\tau ,\zeta ,\bar{\zeta})$, and then eliminating $\tau ,$
using $\tau =T(u,\zeta ,\bar{\zeta})$.

Using the known properties of the tensorial spin-$s$ spherical harmonics and
the $\eth $-operator \cite{Harmonics}, it follows that \textit{regular}
solutions to (\ref{GCE}) are of the form:%
\[
u=z^{a}\hat{l}_{a}=\frac{\sqrt{2}}{2}z^{0}-\frac{1}{2}z^{i}Y_{1i}^{0}, 
\]%
where the $z^{a}$ are four arbitrary complex constants. \ This, for each
choice of the the four constants, gives only one complex cross section of $%
\mathfrak{I}_{\mathbb{C}}^{+}$. The entire family of cut functions may be
obtained by letting $z^{a}$ become $\tau $ dependent, so that the four
arbitrary parameters become an arbitrary complex world-line, $z^{a}=\xi
^{a}(\tau )$:%
\begin{eqnarray}
u &=&G(\tau ,\zeta ,\bar{\zeta})=\xi ^{a}(\tau )\hat{l}_{a}(\zeta ,\bar{\zeta%
})  \label{GCF} \\
&=&\frac{\sqrt{2}}{2}\xi ^{0}(\tau )-\frac{\xi ^{i}(\tau )}{2}%
Y_{1i}^{0}(\zeta ,\bar{\zeta}).  \nonumber
\end{eqnarray}

It thus follows that the function $L$ is given parametrically by%
\begin{eqnarray}
u &=&\xi ^{a}(\tau )\hat{l}_{a}(\zeta ,\bar{\zeta})  \label{L1} \\
L(u,\zeta ,\bar{\zeta}) &=&\xi ^{a}(\tau )\hat{m}_{a}(\zeta ,\bar{\zeta}%
)=\xi ^{i}(\tau )Y_{1i}^{1}(\zeta ,\bar{\zeta}).  \label{L2}
\end{eqnarray}

By a regular solution we mean that the function $L(u,\zeta ,\bar{\zeta})$
should remain finite over the $(\zeta ,\bar{\zeta})$ sphere. Geometrically
this means that all rays of the NGC intersect $\mathfrak{I}^{+}$ i.e., none
lie on $\mathfrak{I}^{+}$ itself.

Before this parametric description of $L(u,\zeta ,\bar{\zeta})$ can be
implemented we must find the restriction on $\tau $ to yield real $u.$ This
is discussed later.

There are three different ways of interpreting the function $L$. \ From (\ref%
{Trans}), we have that $L$ is simply a function which labels NGCs in $%
\mathbb{M}$; by choosing different holomorphic complex functions, we select
different null geodesic congruences in the space-time. \ One can also
interpret\cite{Aronson} $L$ in equation (\ref{Shear2}) as the function
(rotation angle) used to transform from a Bondi tetad frame at $\mathfrak{I}%
^{+}$ to a null tetrad frame where the $l^{\ast a}$ is tangent to a
shear-free NGC; any $L$ which satisfies the PDE (\ref{Shear2}) will result
in such a transformation. \ Finally, the relation (\ref{AngF}), the complex
tangent directions to the cut, allows us to view $L$ as a complex
stereographic angle field on $\mathfrak{I}^{+}$ describing the past null
cone at each point $(u,\zeta ,\bar{\zeta}),$ the direction at which the
outgoing null geodesic intersects with the asymptotic boundary $\mathfrak{I}%
^{+}$ \cite{Review}. \ All three of these views are equally viable, and we
can apply each of them to $L$ as best suits our purposes in what follows.

\textbf{NOTE: }In the context of a discussion of $L(u,\zeta ,\bar{\zeta})$
given parametrically by Eqs.(\ref{L1}) and (\ref{L2}), we must point out,
and be very aware, that there are three other functions that resemble $%
L(u,\zeta ,\bar{\zeta})$ but should not be confused with it. From Eqs.(\ref%
{L1}) and (\ref{L2}) we had

a)

\begin{eqnarray}
u &=&\xi ^{a}(\tau )\hat{l}_{a}(\zeta ,\bar{\zeta})  \label{a.} \\
L(u,\zeta ,\bar{\zeta}) &=&\xi ^{a}(\tau )\hat{m}_{a}(\zeta ,\bar{\zeta}). 
\nonumber
\end{eqnarray}

The similar ones are

b)%
\begin{eqnarray}
\overline{u} &=&\overline{\xi }^{a}(\overline{\tau })\hat{l}_{a}(\zeta ,\bar{%
\zeta}),  \label{b.} \\
\bar{L}(u,\zeta ,\bar{\zeta}) &=&\overline{\xi }^{a}(\overline{\tau })%
\overline{\hat{m}}_{a}(\zeta ,\bar{\zeta}),  \nonumber
\end{eqnarray}%
\qquad \qquad

c)%
\begin{eqnarray}
u &=&\xi ^{a}(\tau )\hat{l}_{a}(\zeta ,\bar{\zeta}),  \label{c.} \\
L^{\ast }(u,\zeta ,\bar{\zeta}) &=&\xi ^{a}(\tau )\overline{\hat{m}}%
_{a}(\zeta ,\bar{\zeta}),  \nonumber
\end{eqnarray}

and

d)%
\begin{eqnarray}
\overline{u} &=&\overline{\xi }^{a}(\overline{\tau })\hat{l}_{a}(\zeta ,\bar{%
\zeta}),  \label{d.} \\
\bar{L}^{\ast }(u,\zeta ,\bar{\zeta}) &=&\overline{\xi }^{a}(\overline{\tau }%
)\hat{m}_{a}(\zeta ,\bar{\zeta}).  \nonumber
\end{eqnarray}

The quantities ($L,\bar{L}$) and ($L^{\ast },\bar{L}^{\ast }$) are complex
conjugate pairs. Only the first pair is basic to our discussion,
nevertheless the second pair does play a role later in Sec.4.

To end this section we emphasize that from the known $L(u,\zeta ,\bar{\zeta}%
) $ and its complex conjugate, when the $\tau $ is taken so that the $u$ is
real, we have the regular shear-free null geodesic given by Eq.(\ref{Trans}),%
\begin{equation}
x^{a}=u\hat{t}^{a}-\bar{L}\hat{m}^{a}-L\overline{\hat{m}}^{a}+(r-r_{0})\hat{l%
}^{a}.  \label{Trans2}
\end{equation}

\subsection{The Optical Parameters}

When describing a NGC in \textit{flat} space-time, two of the Newman-Penrose
spin-coefficients are of particular importance; these are $\rho $ and $%
\sigma $, known as the optical parameters. \ In a null tetrad frame, where
the vector $l$ is tangent to a NGC, these spin-coefficients are defined by 
\cite{Scholarpedia}:%
\begin{eqnarray}
\rho &=&m^{a}\bar{m}^{b}\nabla _{a}l_{b}=\frac{1}{2}\left( -\nabla
_{a}l^{a}+i\func{curl}l^{a}\right) ,  \label{Div} \\
\func{curl}l^{a} &=&\sqrt{\nabla _{\lbrack a}l_{b]}\nabla ^{a}l^{b}}; 
\nonumber
\end{eqnarray}%
\begin{equation}
\sigma =m^{a}m^{b}\nabla _{a}l_{b}.  \label{Shear}
\end{equation}

The parameter $\rho $ is referred to as the complex divergence (often simply
as the divergence) and $\sigma $ is the complex shear of the $l$ null
geodesic congruence. \ The "radial" behavior of the optical equations (i.e.,
their $r$-dependence) is governed by the coupled set of equations known as
the Sachs optical equations:%
\begin{eqnarray}
\frac{\partial \rho }{\partial r} &=&\rho ^{2}+\sigma \bar{\sigma},
\label{OEs} \\
\frac{\partial \sigma }{\partial r} &=&2\rho \sigma .  \nonumber
\end{eqnarray}

The optical equations can be written in the form of a matrix Riccati
equation:%
\begin{equation}
DP=P^{2},  \label{RicE}
\end{equation}%
where the differential operator is $D\equiv \partial _{r}$, and the matrix $%
P $ is:%
\begin{equation}
P=\left( 
\begin{array}{cc}
\rho , & \sigma \\ 
\bar{\sigma}, & \rho%
\end{array}%
\right) .  \label{Mat}
\end{equation}

Integrating Eq.(\ref{RicE}) and fixing the affine parameter origin $r_{0}$
as:%
\begin{equation}
r_{0}=-\frac{1}{2}\left( \eth \bar{L}+L\bar{L}^{\cdot }+\bar{\eth }L+\bar{L}%
L^{\cdot }\right) ,  \label{Origin}
\end{equation}
yields the flat-space solutions \cite{Mink,Review}:

\begin{eqnarray}
\rho &=&\frac{i\Sigma -r}{r^{2}+\Sigma ^{2}-\sigma ^{0}\bar{\sigma}^{0}},
\label{rho1} \\
\sigma &=&\frac{\sigma ^{0}}{r^{2}+\Sigma ^{2}-\sigma ^{0}\bar{\sigma}^{0}},
\label{sig1}
\end{eqnarray}%
where $\Sigma =\Sigma (u,\zeta ,\bar{\zeta})\in \mathbb{R}$ and $\sigma
^{0}=\sigma ^{0}(u,\zeta ,\bar{\zeta})\in \mathbb{C}$ are functions of
integration, called the twist and asymptotic shear of the NGC respectively.
\ It can be shown, via Eq.(\ref{Trans}), that the twist $\Sigma $ is related
to the arbitrary complex function $L$ which selects the NGC in equation (\ref%
{Trans}) by \cite{Mink}%
\begin{equation}
2i\Sigma =\eth \bar{L}+L\bar{L}^{\cdot }-\bar{\eth }L-\bar{L}L^{\cdot }.
\label{Twist1}
\end{equation}

As we are interested in shear-free NGCs in $\mathbb{M}$, we consider those
solutions to (\ref{OEs}) for which $\sigma ^{0}=0$; automatically we see
that all such NGCs are everywhere shear-free:%
\begin{equation}
\rho =-\frac{1}{r+i\Sigma },\ \ \sigma =0.  \label{SFOP}
\end{equation}%
The twist $\Sigma $ is essentially the imaginary part of the complex
divergence $\rho $ of a shear-free NGC in $\mathbb{M}$. \ The caustics of
such a NGC (those points at which $\rho \rightarrow \infty $) are hence
given by:%
\[
r=0,\ \ \Sigma (u,\zeta ,\bar{\zeta})=0, 
\]%
and are on occasion referred to as the `source' of the NGC.

From equations (\ref{AngF}), (\ref{L1}), and (\ref{Twist1}) it should be
noted that if the world-line $\xi ^{a}$ generating a shear-free NGC is taken
to be \textit{real}, then the twist of the congruence vanishes ($\Sigma =0$%
). \ It is in this sense that we may think of the twist of a shear-free NGC
as a measure of how far into $\mathbb{M}_{\mathbb{C}}$ the associated
complex world-line is displaced. \ If we interpret the complex world-line as
the `source' of the congruence (see Sec.4), then twist-free and shear-free
NGCs have their source in $\mathbb{M}$, (a real world-line), while twisting
shear-free NGCs appear (to an observer on $\mathfrak{I}^{+}$) to have their
source in $\mathbb{M}_{\mathbb{C}}$.

\subsection{Real Cuts from Complex Good Cuts}

It can be shown by a limiting process (taking $r\rightarrow \infty $)\cite%
{WL} that the intersection of the complex light-cone from points on the
complex world-line, $\xi ^{a}(\tau )$, with the complex $\mathfrak{I}_{%
\mathbb{C}}^{+}$ is described by the complex cut function, Eq.(\ref{GCF}), 
\begin{equation}
u=G(\tau ,\zeta ,\bar{\zeta})=\frac{\sqrt{2}}{2}\xi ^{0}(\tau )-\frac{\xi
^{i}(\tau )}{2}Y_{1i}^{0}.  \label{GCF2}
\end{equation}

Our immediate task is to find the real values of $u,$ i.e., the intersection
of the complex cone with real $\mathfrak{I}^{+}.$ This entails finding
restrictions on the parameter $\tau .$ This\cite{Review} involves first
splitting $\tau $ into its real and imaginary parts:%
\begin{equation}
\tau =s+i\lambda .  \label{R&I1}
\end{equation}%
The cut function can be written as a function of two real variables $%
(s,\lambda )$ instead of the single complex $\tau $ and then decomposed as:%
\begin{equation}
u=G(s+i\lambda ,\zeta ,\bar{\zeta})=G_{R}(s,\lambda ,\zeta ,\bar{\zeta}%
)+iG_{I}(s,\lambda ,\zeta ,\bar{\zeta}),  \label{R&I2}
\end{equation}%
where the real and imaginary parts of the cut function can be found by%
\begin{eqnarray}
G_{R}(s,\lambda ,\zeta ,\bar{\zeta}) &=&\frac{1}{2}\left[ G(s+i\lambda
,\zeta ,\bar{\zeta})+\overline{G(s+i\lambda ,\zeta ,\bar{\zeta})}\right] ,
\label{GR} \\
G_{I}(s,\lambda ,\zeta ,\bar{\zeta}) &=&\frac{1}{2}\left[ G(s+i\lambda
,\zeta ,\bar{\zeta})-\overline{G(s+i\lambda ,\zeta ,\bar{\zeta})}\right] .
\label{GI}
\end{eqnarray}

The real values of $u$ are then found by setting

\begin{equation}
G_{I}(s,\lambda ,\zeta ,\bar{\zeta})=0.  \label{Im}
\end{equation}

\noindent Solving equation (\ref{Im}) for $\lambda $ then yields%
\begin{equation}
\lambda =\Lambda (s,\zeta ,\bar{\zeta}).  \label{Im2}
\end{equation}%
Combining Eqs.(\ref{Im2}), (\ref{R&I1}) and (\ref{R&I2}) gives us a
real-valued cut function $G:\mathbb{R}\times S^{2}\rightarrow \mathbb{R}$ of
the form%
\begin{equation}
u=G(\tau ^{(\mathrm{R})},\zeta ,\bar{\zeta})\equiv G_{R}(s,\Lambda ,\zeta ,%
\bar{\zeta}),  \label{RGCF}
\end{equation}%
with%
\begin{equation}
\tau ^{(\mathrm{R})}\equiv s+i\Lambda (s,\zeta ,\bar{\zeta}).  \label{R&I3}
\end{equation}

Care must be taken as to when $\tau ^{(\mathrm{R})}$ is inserted into the
complex $G(\tau ,\zeta ,\bar{\zeta})$. \ For instance, the function $L$ is
calculated by first applying $\eth _{(\tau )}$ to the complex $G(\tau ,\zeta
,\bar{\zeta})$ \textit{and then} afterwards inserting $\tau ^{(\mathrm{R})}$:%
\[
L(u,\zeta ,\bar{\zeta})=\left( \eth _{(\tau )}G\right) |_{\tau =\tau ^{(%
\mathrm{R})}}. 
\]%
Taking $L=\eth G_{R}$ would be incorrect, since the additional angular
dependence in $G_{R}$ coming from $\Lambda (s,\zeta ,\bar{\zeta})$ would
yield a different result upon application of the $\eth $-operator. \ In the
next section, we will explore some more consequences of this additional
angular dependence.

\section{Real Geometric Structures from the Complex World-Lines}

In this section, we investigate some of the real affects coming from the
restriction of the complex cuts to real $\mathfrak{I}^{+}$. \ We begin by
considering the differences between the real cuts generated by a complex
world-line and those generated by a real world-line in real $\mathbb{M}$. We
then discuss the analytic description of the shear-free NGCs using the
restriction of the range of $\tau $ to $\tau ^{(\mathrm{R})}=s+i\Lambda
(s,\zeta ,\bar{\zeta}).$ The section concludes with a discussion of the
effects of these results on the associated CR structure induced on $%
\mathfrak{I}^{+}$ by the shear-free NGCs.

\subsection{Real World-lines and Complex World-lines}

Suppose that for our arbitrary world-line generating a shear-free NGC, we
choose $\xi _{R}^{a}(s)\in \mathbb{M}$; that is, we take a real world-line
parametrized by the real variable $s$. \ As pointed out earlier, such a NGC
is twist free ($\Sigma =0$). Regular NGCs generated by this method are just
the null geodesics of the real light-cones with apex on the real
world-line.\ The cut function generated this way is given as%
\begin{equation}
u=G(s,\zeta ,\bar{\zeta})=\frac{\xi _{R}^{0}(s)}{\sqrt{2}}-\frac{\xi
_{R}^{i}(s)}{2}Y_{1i}^{0}.  \label{GCF3}
\end{equation}%
The real parameter $s$ labels the real "cuts" of $\mathfrak{I}^{+}$. \ 

It is easy to see that the constant $s$ cross-sections of $\mathfrak{I}^{+}$
are (undistorted) 2-spheres since they involve only the $l=0,1$ harmonics.
The question is, how does this compare with the cross-sections produced by a
general complex world-line?

For an arbitrary world-line $\xi ^{a}(\tau )\in \mathbb{M}_{\mathbb{C}}$, we
have the (complex) cut function from Eq.(\ref{GCF}):%
\[
G(\tau ,\zeta ,\bar{\zeta})=\frac{\xi ^{0}(\tau )}{\sqrt{2}}-\frac{\xi
^{i}(\tau )}{2}Y_{1i}^{0}. 
\]%
From the previous section, this cut function can be restricted to the real $%
\mathfrak{I}^{+}$ by replacing $\tau $ with $\tau ^{(\mathrm{R})}$, giving
the real cut function:%
\begin{equation}
u=G(\tau ^{(\mathrm{R})},\zeta ,\bar{\zeta})=\frac{\xi ^{0}(s+i\Lambda
(s,\zeta ,\bar{\zeta}))}{\sqrt{2}}-\frac{\xi ^{i}(s+i\Lambda (s,\zeta ,\bar{%
\zeta}))}{2}Y_{1i}^{0}.  \label{RGCF2}
\end{equation}%
Once again, we see that the one-parameter family of slicings of $\mathfrak{I}%
^{+}$ are labeled by the real parameter $s$. \ However, for a fixed $s$, the
cross-section of $\mathfrak{I}^{+}$ we obtain are not in general just $S^{2}$%
. \ This follows because of the additional angular dependence of the
world-line coming from the $\Lambda (s,\zeta ,\bar{\zeta})$ in $\tau ^{(%
\mathrm{R})}$. \ The $\Lambda (s,\zeta ,\bar{\zeta})$ has, in general, an
arbitrary spherical harmonic decomposition, so that the spherical harmonic
expansion of $G(s+i\Lambda (s,\zeta ,\bar{\zeta}),\zeta ,\bar{\zeta})$ will
be distorted from just the $l=0,1$ harmonics of the real world-lines.

We thus have that a major difference between non-twisting and twisting
shear-free NGCs in Minkowski space is that the former produce spherical
slicings of $\mathfrak{I}^{+}$, while the latter have distorted sphere
cross-sections.\ The key ingredient for this distinction is the angular
dependence of $\Lambda (s,\zeta ,\bar{\zeta})$ which is needed when we force
the complex cut function to take on real values.

\subsection{The CR Structures and Levi Forms}

It is known that the $\mathfrak{I}^{+}$ of Minkowski space (or even
asymptotically flat space-times) has a realizable CR structure associated
with each choice of a shear-free NGC (or equivalently with the choice of a
complex world-line generating the congruences)\cite{ScriCR,Review}. \ As a
real 3-dimensional manifold, $\mathfrak{I}^{+}$ is said to be a CR manifold
(or have an associated CR structure) if there is a real 1-form $\mathfrak{L}$
and complex 1-form $\mathfrak{M}$ defined up to the CR gauge transformations%
\begin{eqnarray}
\mathfrak{L} &\rightarrow &a\mathfrak{L},  \label{CR1} \\
\mathfrak{M} &\rightarrow &f\mathfrak{M}+g\mathfrak{L},  \nonumber
\end{eqnarray}%
where $(a,f,g)$ are functions on $\mathfrak{I}^{+}$, with $a$ real and
non-vanishing, $f$ complex and non-vanishing, and $g$ complex. \ These
1-forms must also be linearly independent in the sense that%
\begin{equation}
\mathfrak{L}\wedge \mathfrak{M}\wedge \mathfrak{\bar{M}}\neq 0.  \label{CR2}
\end{equation}

On the $\mathfrak{I}^{+}$ of Minkowski space, using the complex function $%
L(u,\zeta ,\bar{\zeta})$ describing a shear-free NGC, these 1-forms (modulo
gauge freedom) are:%
\begin{eqnarray}
\mathfrak{L} &=&du-\frac{L}{1+\zeta \bar{\zeta}}d\zeta -\frac{\bar{L}}{%
1+\zeta \bar{\zeta}}d\bar{\zeta},  \label{CR3} \\
\mathfrak{M} &=&\frac{d\overline{{\zeta }}}{1+\zeta \bar{\zeta}},  \nonumber
\end{eqnarray}%
with the associated vector duals%
\begin{eqnarray}
l &=&\frac{\partial }{\partial u},  \label{CR4} \\
m &=&(1+\zeta \bar{\zeta})\frac{\partial }{\partial \zeta }+L\frac{\partial 
}{\partial u_{B}}.  \nonumber
\end{eqnarray}%
This CR structure is said to be realizable if we can find an embedding of
the form $\iota :\mathfrak{I}^{+}\rightarrow \mathbb{C}^{2}$, where the
function $\iota (u,\zeta ,\bar{\zeta})$ is given by the two linearly
independent solutions $K_{1}$ and $K_{2}$ of the CR equation:%
\begin{equation}
mK_{i}=\eth K_{i}+LK_{i}^{\cdot }=0  \label{CR5}
\end{equation}%
The first solution is simply $K_{1}(u,\zeta ,\bar{\zeta})=\bar{\zeta}$, with
the second given by $K_{2}(u,\zeta ,\bar{\zeta})=T(u,\zeta ,\bar{\zeta})$,
the complex potential function introduced earlier. \ It is in this sense
that (\ref{CReq}) is a CR equation, and $T$ a corresponding CR function. \
So we have that in $\mathbb{M}$, shear-free NGCs generate realizable CR
structures on $\mathfrak{I}^{+}$ of the form%
\[
\iota (u,\zeta ,\bar{\zeta})=(\bar{\zeta},\tau ), 
\]%
and that each choice of complex world-line $\xi ^{a}$ induces a different
corresponding CR structure.

Every CR manifold of this type is endowed with a Hermitian 2-form (the Levi
form), which encodes information about the pseudo-convexity of the manifold 
\cite{CR}. \ For the CR structure on $\mathfrak{I}^{+}$, we can calculate
the Levi form as%
\begin{equation}
h=-2i\left[ d\mathfrak{L}\lfloor \left( m\otimes \bar{m}\right) \right] ,
\label{Levi2}
\end{equation}%
where $\mathfrak{L}$ is taken from (\ref{CR3}), $m$ and $\bar{m}$ are taken
from (\ref{CR4}), and "$\lfloor $" stands for contraction. \ 

Now, writing $P\equiv 1+\zeta \bar{\zeta}$, we have that:%
\begin{eqnarray}
d\mathfrak{L} &=&\frac{L^{\cdot }}{P}d\zeta \wedge du-\frac{\bar{L}^{\cdot }%
}{P}du\wedge d\bar{\zeta}  \label{Levi3} \\
&&+\left[ \frac{\partial }{\partial \bar{\zeta}}\left( \frac{L}{P}\right) -%
\frac{\partial }{\partial \zeta }\left( \frac{\bar{L}}{P}\right) \right]
d\zeta \wedge d\bar{\zeta}.  \nonumber
\end{eqnarray}%
Denoting components as $d\mathfrak{L}\equiv \mathfrak{L}_{ab}$ and $m\otimes 
\bar{m}\equiv M^{ab}$ respectively, it is then a simple calculation to see
that%
\[
\mathfrak{L}_{ab}M^{ab}=\bar{\eth }L+\bar{L}L^{\cdot }-\eth \bar{L}-L\bar{L}%
^{\cdot }, 
\]%
and it thus follows that the Levi form for the shear-free CR structure on $%
\mathfrak{I}^{+}$ is%
\begin{equation}
h=-2i(\bar{\eth }L+\bar{L}L^{\cdot }-\eth \bar{L}-L\bar{L}^{\cdot }).
\label{Levi4}
\end{equation}%
From equation (\ref{Twist1}) it is we see that $h$ is proportional to $%
\Sigma $, the twist of the NGC. \ Combining this with the discussion of the
previous subsection yields the following result:

We see immediately that any CR structure generated by a real world-line $\xi
^{a}\in \mathbb{M}$ (i.e., twist-free) induces a Levi-flat CR structure on $%
\mathfrak{I}^{+}$.

\subsection{Parametrically Describing the Shear-free NGC}

Earlier, in Eq.(\ref{Trans}), we gave the general expression for any
Minkowski space NGC where the $L(u,\zeta ,\bar{\zeta})$ was an arbitrary
function. If the congruence was to be shear free $L(u,\zeta ,\bar{\zeta})$
had to satisfy the shear free condition, Eq.(\ref{Shear2}). From the
parametric relations;

\begin{eqnarray}
L(u,\zeta ,\bar{\zeta}) &=&\eth _{(\tau )}G(\tau ,\zeta ,\bar{\zeta})|_{\tau
=s+i\Lambda (s,\zeta ,\bar{\zeta})},  \label{Para1} \\
u &=&G(s+i\Lambda (s,\zeta ,\bar{\zeta}),\zeta ,\bar{\zeta}),  \label{Para 2}
\\
&=&\xi ^{a}(s+i\Lambda (s,\zeta ,\bar{\zeta}))l_{a}
\end{eqnarray}
we have the parametric description of the shear free NGCs in terms of ($%
u,\zeta ,\bar{\zeta}$)

\[
x^{a}=u\hat{t}^{a}-\bar{L}\hat{m}^{a}-L\overline{\hat{m}}^{a}+(r-r_{0})\hat{l%
}^{a}. 
\]

Equivalently, we can express the Minkowski space coordinates in terms of the
real parameter $s$, which labels the leaves of the good cut foliation of $%
\mathfrak{I}^{+}$. \ To do this, we simply take the $L$ before replacing the 
$\tau $ by $u$ from (\ref{L1}), i.e.,%
\[
L(u(\tau ),\zeta ,\bar{\zeta})=\xi ^{i}(\tau )Y_{1i}^{1}(\zeta ,\bar{\zeta}%
), 
\]%
and replace the $u$ in (\ref{Trans}) by $G(\tau ,\zeta ,\bar{\zeta})$. \ The
resulting expression when restricted to $\tau ^{(\mathrm{R})}=s+i\Lambda
(s,\zeta ,\bar{\zeta})$ yields the explicit description of any flat space
shear free NGC in terms of ($s,\zeta ,\bar{\zeta}$).%
\begin{eqnarray}
x^{a} &=&\xi ^{b}(\tau ^{(\mathrm{R})})l_{b}\hat{t}^{a}-\bar{L}(\overline{%
\tau }^{(\mathrm{R})}))\hat{m}^{a}-L(\tau ^{(\mathrm{R})})\overline{\hat{m}}%
^{a}+(r-r_{0})\hat{l}^{a}  \label{Para2} \\
&=&\left( \frac{\xi ^{0}(\tau ^{(\mathrm{R})})}{\sqrt{2}}-\frac{\xi
^{i}(\tau ^{(\mathrm{R})})}{2}Y_{1i}^{0}\right) \hat{t}^{a}-\overline{\xi }%
^{i}(\overline{\tau }^{(\mathrm{R})})Y_{1i}^{-1}\hat{m}^{a}-\xi ^{i}(\tau ^{(%
\mathrm{R})})Y_{1i}^{1}\overline{\hat{m}}^{a}+(r-r_{0})\hat{l}^{a}. 
\nonumber
\end{eqnarray}

Care must be taken here when calculating $r_{0}$, which involves terms of
the form $\bar{\eth }L$. \ Here, the $\eth $-operator should first be
applied to $L(u(\tau ),\zeta ,\bar{\zeta})\,$\ and only then insert the $%
\tau ^{(\mathrm{R})}$; this is again because of the added angular dependence
coming from $\Lambda (s,\zeta ,\bar{\zeta})$. \ In these new coordinates, $%
(s,\zeta ,\bar{\zeta})$ label the geodesics of the congruence by naming them
for the spot where they intersect $\mathfrak{I}^{+}.$ The $r$ is the affine
parameter along each geodesic.

\section{Shear-Free Congruence Directly from Complex World-line}

Our principle result is the demonstration that the real shear-free NGC can
be (easily) found by mapping the complex world-line into the real Minkowski
space. \ 

\ \textbf{Theorem}: There exists a mapping from the complex world-line $\xi
^{a}(\tau )\in \mathbb{M}_{\mathbb{C}}$ to the real shear-free NGC in $%
\mathbb{M}$ given by two complex null displacements.

Beginning with the world line, $\xi ^{a}(\tau ),$ written in terms of its
components ($\xi ^{b}l_{b},\xi ^{b}n_{b},\xi ^{b}m_{b},\xi ^{b}\overline{m}%
_{b}$) as%
\begin{equation}
\xi ^{a}(\tau )=\xi ^{b}(\tau )\hat{l}_{b}\hat{n}^{a}+\xi ^{b}(\tau )%
\widehat{n}_{b}\hat{l}^{a}-\xi ^{b}(\tau )\hat{m}_{b}\overline{\hat{m}}%
^{a}-\xi ^{b}(\tau )\overline{\hat{m}}_{b}\hat{m}^{a}  \label{1.}
\end{equation}%
we replace the $n^{a}$ by%
\begin{eqnarray*}
\hat{n}^{a} &=&\sqrt{2}t^{a}-\hat{l}^{a} \\
t^{a} &=&\delta _{0}^{a}
\end{eqnarray*}%
yielding%
\begin{equation}
\xi ^{a}(\tau )=\sqrt{2}\xi ^{b}(\tau )\hat{l}_{b}t^{a}+\sqrt{2}\xi
^{b}(\tau )t_{b}\hat{l}^{a}-2\xi ^{b}(\tau )\hat{l}_{b}\widehat{l}^{a}-\xi
^{b}(\tau )\hat{m}_{b}\overline{\hat{m}}^{a}-\xi ^{b}(\tau )\overline{\hat{m}%
}_{b}\hat{m}^{a}.  \label{2.}
\end{equation}%
Remembering, from Eqs.(\ref{L1}) and (\ref{c.}), that%
\begin{eqnarray*}
u &=&\xi ^{b}(\tau )\widehat{l}_{b} \\
L^{\ast }(u,\zeta ,\bar{\zeta}) &=&\xi ^{a}(\tau )\overline{\hat{m}}%
_{a}(\zeta ,\bar{\zeta}),
\end{eqnarray*}%
Eq.(\ref{2.}) becomes

\begin{equation}
\xi ^{a}(\tau )=\sqrt{2}ut^{a}+\sqrt{2}\xi ^{0}(\tau )\hat{l}^{a}-2u\hat{l}%
^{a}-L\overline{\hat{m}}^{a}-L^{\ast }\hat{m}^{a},
\end{equation}%
or%
\begin{equation}
\xi ^{a}(\tau )+2u\hat{l}^{a}-\sqrt{2}\xi ^{0}(\tau )\hat{l}^{a}+L^{\ast }%
\hat{m}^{a}=\sqrt{2}ut^{a}-L\overline{\hat{m}}^{a}.  \label{3.}
\end{equation}%
Finally subtracting $\overline{L}\hat{m}^{a}$ and adding $(r-r_{0})\hat{l}%
^{a}$ to both sides of Eq.(\ref{3.}) we obtain%
\begin{eqnarray}
x^{a} &=&\xi ^{a}(\tau )+(2u-\sqrt{2}\xi ^{0}(\tau ))\hat{l}^{a}+(L^{\ast }-%
\overline{L})\hat{m}^{a}+(r-r_{0})\hat{l}^{a}  \label{4.} \\
&=&\sqrt{2}ut^{a}-L\overline{\hat{m}}^{a}-\overline{L}\hat{m}^{a}+(r-r_{0})%
\hat{l}^{a}.  \nonumber
\end{eqnarray}%
To complete our task we now restrict the values of $\tau $ to those that
produce a real $u,$ namely 
\[
\tau \rightarrow \tau ^{(\mathrm{R})}=s+i\Lambda (s,\zeta ,\bar{\zeta}). 
\]

We see that by adding several null vectors proportional to $\hat{m}^{a}$ and 
$\hat{l}^{a}$ directly to the complex world-line $\xi ^{a}(\tau ),$ we
obtain a mapping of the world-line directly to the real shear-free NGC, Eq.(%
\ref{Para2}).

We thus have the explicit relationship of the complex world-line to the
shear-free NGC.

\section{Caustic Geometry}

As mentioned in Section 2, the caustic set, where the complex divergence $%
\rho $ of the NGC diverges, can be interpreted as the `source' of the NGC
itself. \ In this section, we will derive the coordinate expression of the
caustics of the NGC in Minkowski space. In doing this, we assume certain
generic behavior regarding the invertibility of several of the functions
which appear. From this we see that the caustic appears as a closed string
or loop that evolves with time in the Minkowski space. \ The non-generic
special cases appear to have the same caustic structure.

\subsection{Generic World-line Caustics}

Recall that for a shear-free NGC in $\mathbb{M}$, the complex divergence has
the form:%
\begin{equation}
\rho =-\frac{1}{r+i\Sigma },  \label{C1}
\end{equation}%
and caustics occur (in the geodesic coordinates) where $\rho \rightarrow
\infty $. \ Since both $r$ and $\Sigma $ are real, it follows immediately
that the caustic set is given by the relationships%
\begin{equation}
r=0,\ \ \Sigma (u,\zeta ,\bar{\zeta})=0.  \label{C2}
\end{equation}%
Putting $r=0$ into the Minkowski space coordinate formula (\ref{Trans}), we
get:%
\begin{equation}
x^{a}=u\hat{t}^{a}-\bar{L}\hat{m}^{a}-L\overline{\hat{m}}^{a}-r_{0}\hat{l}%
^{a},  \label{Caus}
\end{equation}%
where $r_{0}=-\frac{1}{2}\left( \eth \bar{L}+L\bar{L}^{\cdot }+\bar{\eth }L+%
\bar{L}L^{\cdot }\right) $ from (\ref{Origin}). Everything on the right hand
side is a function of $(u,\zeta ,\bar{\zeta})$. \ For a fixed time in
Minkowski coordinates, say $x^{0}=t$, we then have from Eqs.(\ref{Caus}), (%
\ref{Stet}) and (\ref{t})%
\begin{equation}
t=x^{0}(u,\zeta ,\bar{\zeta})\equiv \sqrt{2}u-\frac{\sqrt{2}}{2}%
r_{0}(u,\zeta ,\bar{\zeta}).  \label{C3}
\end{equation}%
By the generic assumption Eq.(\ref{C3}) can be inverted to give:%
\begin{equation}
u=H(t,\zeta ,\bar{\zeta}).  \label{C4}
\end{equation}

Also assuming that%
\begin{equation}
\Sigma (u,\zeta ,\bar{\zeta})=-\frac{i}{2}\left( \eth \bar{L}+L\bar{L}%
^{\cdot }-\bar{\eth }L-\bar{L}L^{\cdot }\right) =0  \label{C5}
\end{equation}%
can be solved for $u$, yields 
\begin{equation}
u=F(\zeta ,\bar{\zeta}).  \label{C6}
\end{equation}%
Eliminating $u$ from (\ref{C4}) and (\ref{C6}) gives%
\begin{equation}
H(t,\zeta ,\bar{\zeta})-F(\zeta ,\bar{\zeta})\equiv K(t,\zeta ,\bar{\zeta}%
)=0.  \label{C7}
\end{equation}

Writing $\zeta =x+iy,$ $\overline{\zeta }=x-iy$ and assuming that (\ref{C7})
can indeed be solved for $y$, we finally obtain:%
\begin{equation}
y=Z(t,x).  \label{C8}
\end{equation}%
This allows us to combine (\ref{C4}), (\ref{C6}), and (\ref{C8}) in the
spatial portion of (\ref{Caus}) to get:%
\begin{equation}
x^{i}=X^{i}(t,x).  \label{C9}
\end{equation}%
Note, from the stereographic projection, that the range of $x$ is between
plus and minus infinity with plus and minus infinity being identified with
each other. \ This means that for our generic world-line the NGC's caustics
at a fixed Minkowski time $t$ will be an $S^{1}$ worth of spatial points.
The space-time caustics of the generic shear-free NGCs in $\mathbb{M}$ are
thus composed of evolving closed strings, i.e., tube-like structures, where
movement up or down the tube is in Minkowski time $t$.

Though it is hard to analyze all possible non-generic cases, nevertheless it
appears as if the above caustic structure is preserved in all cases. For
example if the twist $\Sigma $ does not depend on $u$ it is easy to see that
we have the same closed curve at any instant of time. The Kerr congruence is
a special case of this.

\section{Conclusion}

We have found a surprising real effect in Minkowski space-time that
originates with an arbitrary complex analytic world-line in complex
Minkowski space, namely that it generates a regular shear-free NGC and that
all such congruences have their origin in such a complex line. One first
`sees' the complex line by the intersection of the complex light-cone with
real null infinity, $\mathfrak{I}^{+},$ where the distortions of the cut
come from the displacement of the world-line from the real into the complex.
Likewise the twist of the NGC is a direct measure of this displacement. The
most complete relationship between the complex world-line and the NGC is the
mapping of the world-line via two complex null displacements directly into
the congruence itself.

Although we have restricted our discussion to flat space-time, several of
the ideas developed here do generalize to asymptotically flat space-times. \
In such space-times, one still finds that asymptotically shear-free NGCs
have an associated complex world-line `living' in an auxiliary Minkowski
space. The same restriction procedure is performed on $\tau $ to produce
real cuts \cite{Review}. \ The asymptotic twist is again a measure of the
displacement of the line into the complex.

\section{Acknowledgment}

\qquad We thank Roger Penrose for several very useful discussions. TMA
acknowledges support from the University of Oxford Clarendon Fellowship and
Balliol College.

\end{document}